\documentclass[aps,prl,twocolumn,superscriptaddress,showpacs,reprint]{revtex4-1}

\usepackage{times}
\usepackage{mathptmx}
\usepackage{amsmath}
\usepackage{amsfonts}
\usepackage{amssymb}
\usepackage{graphicx}
\usepackage{hyperref}

\def\addSouth{School of Mathematics, University of Southampton, Southampton SO17 1BJ, United Kingdom}
\def\addCbridge{Institute of Astronomy, Madingley Road, Cambridge, CB3 0HA, United Kingdom}
\def\addYuk{Yukawa Institute for Theoretical Physics, Kyoto University, Kyoto 606-8502, Japan}

\begin{document}

\title{Evolution of inspiral orbits around a Schwarzschild black hole}

\author{Niels Warburton}
\affiliation{\addSouth}

\author{Sarp Akcay}
\affiliation{\addSouth}

\author{Leor Barack}
\affiliation{\addSouth}

\author{Jonathan R.\ Gair}
\affiliation{\addCbridge}

\author{Norichika Sago}
\affiliation{\addYuk}

\begin{abstract}
We present results from calculations of the orbital evolution in eccentric binaries of nonrotating black holes with extreme mass-ratios. Our inspiral model is based on the method of osculating geodesics, and is the first to incorporate the full gravitational self-force (GSF) effect, including conservative corrections. The GSF information is encapsulated in an analytic interpolation formula based on numerical GSF data for over a thousand sample geodesic orbits. We assess the importance of including conservative GSF corrections in waveform models for gravitational-wave searches.
\end{abstract}


\maketitle

{\it Introduction.} The relativistic two-body problem for binaries of greatly different masses is of both theoretical importance and astrophysical relevance. When the mass ratio is extreme, deviation from geodesic motion can be described in terms of an effective gravitational self-force (GSF) arising from the interaction of the small object with its own spacetime perturbation. The fundamental problem of regularizing the gravitational self-interaction in curved spacetime has been studied extensively since the late 1990s, and is now rigorously solved at the first post-geodesic order (i.e., at linear order in the small mass ratio $\eta$) \cite{Mino:1996nk,Poisson:2003nc,Gralla:2008fg,Pound:2009sm}. This theoretical advance has been strongly motivated by the exciting prospects of observing gravitational waves from inspiralling black hole binaries of small mass ratios. Systems with $\eta= 10^{-2}$--$10^{-3}$ (IMRIs: intermediate-mass-ratio inspirals) could be detected by Advanced LIGO \cite{Brown:2006pj}, and a low-frequency detector in space based on the LISA design will observe many inspiralling systems with $\eta=10^{-4}$--$10^{-7}$ (EMRIs: extreme-mass-ratio inspirals) \cite{Gair:2008bx}. The latter, involving a compact object captured by a massive black hole, are of key importance in gravitational-wave astronomy due to their unique utility as probes of strong-field gravity \cite{Barack:2003fp,AmaroSeoane:2007aw}. To interpret the information encoded in the I/EMRI signals it is crucial to have at hand an accurate model of the orbital evolution driven by the GSF.

The last decade has seen a concentrated effort to develop computational tools for the GSF in black hole spacetimes \cite{Barack:2009ux}. This culminated in 2010 \cite{Barack:2010tm} with the introduction of a code that returns the GSF along any specified (fixed) bound geodesic orbit of the Schwarzschild geometry. Results from this code have already been used to quantify aspects of the conservative post-geodesic dynamics \cite{Barack:2011ed}, and today they provide a unique strong-field benchmark for post-Newtonian calculations \cite{Favata:2010yd}, nonlinear numerical simulations \cite{Tiec:2011bk,Tiec:2011dp}, and Effective One Body theory \cite{Damour:2009sm,Barack:2010ny,Barausse:2011dq}. However, with the I/EMRI problem in mind, it remains an important task to translate the GSF information into inspiral trajectories and gravitational waveforms. This has not been attempted so far. 

Here we report an important milestone in the GSF programme: an algorithm and a working code for computing inspiral orbits on a Schwarzschild background, incorporating the full GSF information. The full GSF has a dissipative piece, responsible for the orbital decay, but also a conservative component, which, e.g., modifies the rate of periastron precession. In the {\it radiative approximation} (RA) one ignores the conservative GSF effect and considers only the secular, time-averaged part of the dissipative dynamics. The latter can be computed using global energy-momentum balance considerations, without resorting to the local GSF \cite{Mino:2003yg}. The RA can bring considerable computational saving, so it is important to assess its efficacy, which we do here reliably for the first time. 

There are two approaches to self-forced evolution. In the systematic approach one solves the perturbation equations and the self-forced equations of motion as a coupled set, in a self-consistent manner. This entails incorporating back-reaction corrections in the GSF code itself---a technically challenging task yet to be attempted. The second approach invokes the traditional method of {\it osculating orbits} (also known in Newtonian celestial mechanics as the method of variation of constants). In this approach the inspiral orbit is reconstructed as a smooth sequence of geodesics, each lying tangent to the orbit at a particular moment. This amounts to modelling the true orbit as an evolving geodesic with dynamical orbital elements. Equations governing the forced evolution of the latter in the Schwarzschild case (with an arbitrary forcing agent) were obtained by Pound and Poisson \cite{Pound:2007th}, and Gair {\it et al.}~generalized the formalism to Kerr \cite{Gair:2010iv}. We adopt here the formalism of \cite{Pound:2007th}, and implement it with actual GSF data for the first time.  

Throughout this letter we set $G=c=1$ and use metric signature $({-}{+}{+}{+})$ and Schwarzschild coordinates $\{t,r,\theta,\varphi\}$. $M$ denotes the mass of the background Schwarzschild geometry, and $\mu$ is the mass of the inspiralling object (so $\mu/M=\eta$).

{\it Osculating geodesics.} Bound geodesics of the Schwarzschild geometry can be parametrized by their semilatus rectum $pM$ and eccentricity $e$, defined via $r_{\pm}=pM/(1\mp e)$, where $r=r_{+}$ and $r=r_{-}$ are the apastron and periastron radii, respectively. The geodesic motion of a test particle is described by 
\begin{equation} \label{r}
r=r_g(t;p,e,\chi_0)=\frac{pM}{1+e\cos[\chi(t)-\chi_0]},
\end{equation}
\begin{equation} \label{varphi}
\varphi=\varphi_g(t;p,e,\chi_0)=\int_{\chi(0)}^{\chi(t)} \frac{{p}^{1/2}\, d\chi'}{\sqrt{p-6-2e\cos(\chi'-\chi_0)}}  ,
\end{equation}
where $\chi(t)$ is a monotonically increasing parameter along the orbit, obtained by inverting $t(\chi)=\int_{\chi_0}^{\chi}(dt/d\chi')d\chi'$ with 
\begin{equation}\label{dtdchi}
\frac{dt}{d\chi}=\frac{Mp^2(1+e\cos v)^{-2}}{p-2-2e\cos v}\left[\frac{(p-2)^2-4e^2}{p-6-2e\cos v}\right]^{1/2}.
\end{equation}
Here $v\equiv \chi-\chi_0$, and without loss of generality we assumed the motion takes place in the equatorial plane ($\theta=\pi/2$), and $t(\chi_0)=\varphi(\chi_0)=0$ ($t=0$ is a periastron passage with $\varphi=0$). 

In the osculating geodesics approach, the inspiral motion of a mass particle under the effect of the GSF is described by $r=r_g(t;p(t),e(t),\chi_0(t))$ and $\varphi=\varphi_g(t;p(t),e(t),\chi_0(t))$, where $p(t)$, $e(t)$, $\chi_0(t)$ are {\it osculating elements}. The {\it principal} elements $p$ and $e$ determine the ``shape'' of the orbit; the {\it positional} element $\chi_0$ describes the orientation of the major axis. Both principal and positional elements evolve secularly under the GSF effect, but while the secular evolution of $p$ and $e$ is dissipative, that of $\chi_0$ is conservative (it describes the {\it precession} effect of the GSF). Both principal and positional elements also exhibit quasi-periodic oscillations. 

Given the GSF components $F^{\alpha}(\propto \eta^2)$, the osculating elements evolve according to \cite{Pound:2007th}
\begin{eqnarray}
\dot{p} \!\!\! & = & \!\!\! 2p f_0 f_1 \!\! \left[\! p^{1/2}f_1 f_2 (p\! -\! 3 \! -\! e^2\cos^2 v)M\tilde F^{\varphi}
- e\sin v\, \tilde F^r \! \right]\!\!\! , 
\label{pdot} \\
\dot{e} \!\!\! & = & \!\!\! p^{1/2}f_0 f_2\left[\beta f_3\cos v+e(p^2-10p+12+4e^2)\right] M\tilde F^{\varphi} 
\nonumber\\	
&& + \beta f_0  f_1 \sin v \,\tilde F^r, 
\label{edot} \\
\!\!\!\!\!\!\!\!\!\!
\dot{\chi_0} \!\!\! & = & \!\!\! p^{1/2}e^{-1}f_0 f_2 \sin v \left[(p-6)f_3-4e^3\cos v\right] M\tilde F^{\varphi} 
\nonumber\\	
&& -e^{-1}f_0 f_1\left[(p-6)\cos v+2e\right] \tilde F^r, 
\label{chidot}
\end{eqnarray}
where an overdot denotes $d/dt$, $\tilde F^{\alpha}\equiv \mu^{-1}F^{\alpha}$ is the self-acceleration, 
$f_0= (p-2-2e\cos v)(p-3-e^2)[(p-2)^2-4e^2]^{-1/2}[(p-6)^2-4e^2]^{-1}$,
$f_1= (p-6-2e\cos v)^{1/2}$, 
$f_2= (1+e\cos v)^{-2}$,
$f_3=f_1^2 e\cos v+2(p-3)$ and
$\beta= p-6-2e^2$.
In our implementation, the GSF data will be given in the form $\tilde F^{\alpha}=\tilde F^{\alpha}(\chi-\chi_0;p,e)$, evaluated along geodesics with fixed $p,e,\chi_0$. With these data at hand, Eqs.\ (\ref{pdot})--(\ref{chidot}) [with Eq.\ (\ref{dtdchi})] form a closed set of ordinary differential equations for $\{p(t),e(t),\chi_0(t)\}$. We will solve this set with the initial conditions $\{p(0),e(0),\chi_0(0)\}=\{p_0,e_0,0\}$ for some $p_0,e_0$. The inspiral trajectory will then be described by Eqs.\ (\ref{r}) and (\ref{varphi}) with $p,e,\chi_0$ replaced by the corresponding osculating elements. 

{\it GSF interpolation model.} 
Existing codes do not return the true GSF along the evolving orbit, but an approximation thereof computed along fixed geodesics. The resulting error is very small in the adiabatic regime where the evolution occurs on a timescale much longer than the orbital period $T$. It can be shown \cite{Cutler:1994pb} that the adiabaticity condition $\alpha\equiv \langle|\dot{p}/p|T\rangle \ll 1$ (where $\langle\cdot\rangle$ denotes an average over time $T$) is met so long as $\epsilon\gg \eta^{1/2}$, where $\epsilon\equiv p-6-2e$ is a measure of the proximity to the innermost stable orbit (ISO). Thus, for EMRI-relevant $\eta$ values, the evolution is adiabatic until very close to the ISO. Beyond that point, our GSF model may cease to be useful. 

GSF codes return $F^{\alpha}(\chi)$ for given $p,e,\chi_0$. To express this information in a workable form we devise accurate analytic fits to numerical data obtained using two independent codes: the original, time-domain code of Ref.\ \cite{Barack:2010tm}, 
and a new code based on a frequency-domain treatment of the Lorenz-gauge perturbation equations \cite{Akcay:2010dx,WAB}. 
Both codes take as input the geodesic parameters $p,e$, and return (separately) the dissipative and conservative pieces of the GSF along the geodesic, $F_{\rm diss}^{\alpha}(\chi;p,e)$ and $F_{\rm cons}^{\alpha}(\chi;p,e)$ respectively. The new, frequency-domain, algorithm offers significant computational saving, particularly at low eccentricity. This is a crucial improvement, since GSF calculations are extremely computationally intensive. 
We used our frequency-domain code to compute the GSF for a dense sample of $p,e$ values in the range $0\leq e\leq 0.2$ and $6+2e< p\leq 12$. We tested a subset of the results using our time-domain code. The results described below are based on a sample of $1100$ geodesics, for which the GSF has been computed with fractional accuracy $\lesssim 10^{-4}$. 

To devise an interpolation formula for the numerical data, we observe that the GSF is a periodic function of $v=\chi-\chi_0$ along the geodesic, with $F_{\rm diss}^{\varphi},F_{\rm cons}^{r}$ even in $v$, and $F_{\rm diss}^{r},F_{\rm cons}^{\varphi}$ odd in $v$ \cite{Barack:2009ux}. This suggests the Fourier-like representation 
\begin{equation} \label{fit}
{\cal F}_i= \sum_{n=0}^{\bar n_i}
\sum_{j=0}^{\bar j_i} \sum_{k=0}^{\bar k_i} {a}_{injk}\, e^{n+2j} p^{-k_i-k} {\rm osc}_i(nv)
\end{equation}
$(i=1,\ldots,4)$,
where ${\cal F}_i\equiv \eta^{-2}\{F_{\rm cons}^{r},F_{\rm diss}^{r},MF_{\rm cons}^{\varphi},MF_{\rm diss}^{\varphi}\}$ and  ${\rm osc}_i(nv)=\{\cos nv,\, \sin nv,\, \sin nv,\, \cos nv \}$. We have employed here a simple power series model for the $p,e$ dependence of the Fourier coefficients. For the leading $1/p$ power we take $k_i=\left\{2,\frac{9}{2},4,\frac{11}{2}\right\}$, consistent with the known behavior of the various ${\cal F}_i$'s at large $p$. Each Fourier $n$-mode of ${\cal F}_i$ admits a power series in $e^2$ starting at $e^n$. 
The dimensionless numerical coefficients $a_{injk}$ in Eq.\ (\ref{fit}) are to be determined by fitting to numerical GSF data, with the summation cutoffs $\bar n_i,\bar j_i,\bar k_i$ to be chosen empirically.

We used a standard least-squares algorithm to fit the interpolation formula (\ref{fit}) to the numerical data over the range of $p,e$-values indicated above. For our illustrative computation we sought a fractional accuracy $<10^{-3}$ in each component ${\cal F}_i$ [i.e., we demanded that Eq.\ (\ref{fit}) reproduced all available data to within that accuracy]. We found empirically that this can be achieved with $\bar n_i=6$, $\bar j_i=2$ and $\bar k_i=9$ for each $i$. Thus the procedure fits $7\times 3\times 10=210$ parameters $a_{injk}$ using 1100 data points for each $i$. For lack of space we do not give here the values of the best-fit model parameters $a_{injk}$, but we have made them available online on a dedicated website \cite{webpage} as part of an open-source ``fast GSF calculator''. The package contains a script for computing the GSF quickly based on Eq.\ (\ref{fit}) and a database of $a_{injk}$ coefficients. We intend to update the database regularly as more GSF data (of improved accuracy and greater extent in the $p,e$ space) become available.



\begin{figure}[h!]
	\includegraphics[scale=0.55]{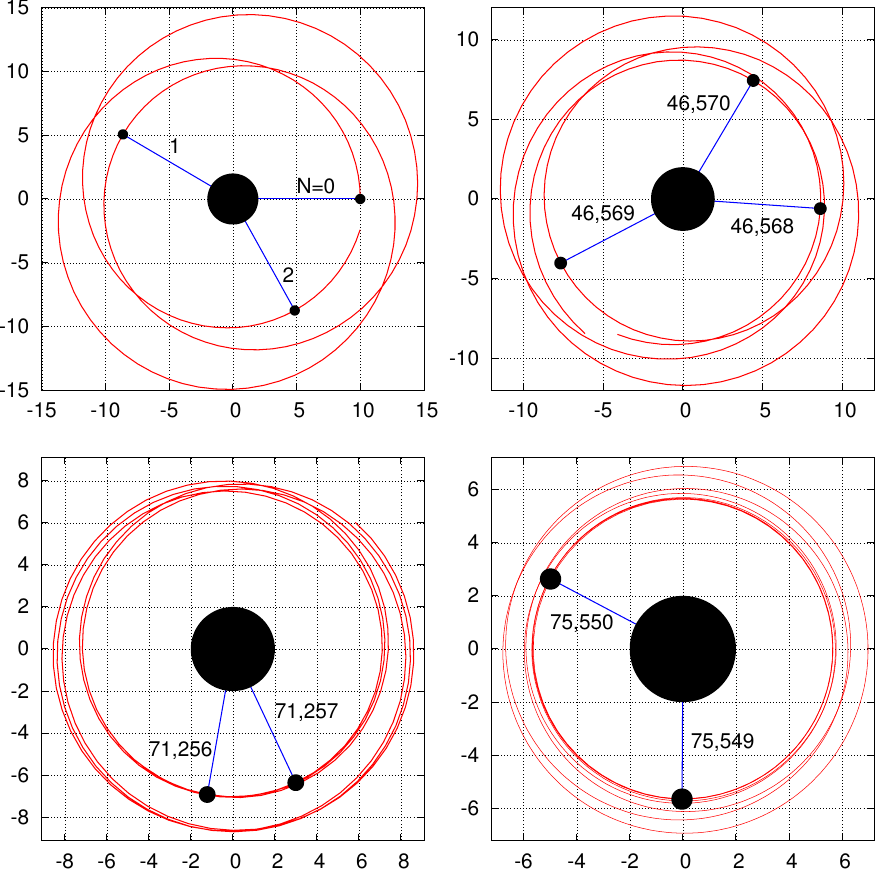}
	\caption{\footnotesize 
	Sample full-GSF inspiral orbit with $\mu=10 M_{\odot}$ and $M=10^6 M_{\odot}$, starting at $(p_0,e_0)=(12,0.2)$. We plot the orbit in the plane of $x=(r/M)\cos\varphi$ and $y=(r/M)\sin\varphi$, showing 4 episodes during the inspiral: the onset of inspiral ($\sim 1443$ days to plunge; top left), 500 days to plunge (top right), 75 days to plunge (bottom left), and the last hour of inspiral (bottom right). The motion is counter-clockwise, and each of the tracks shows one hour of inspiral. The central black hole (but not the orbiter) is drawn to scale. Periastron passages are indicated along with their sequential number, counting from the initial periastron (`$N=0$'). In the last snapshot the orbit completes $\sim 6.7$ revolutions in $\varphi$ between the two periastra shown.
}
	\label{fig:1}
\end{figure}

\begin{figure}[h!]
	\includegraphics[scale=0.72]{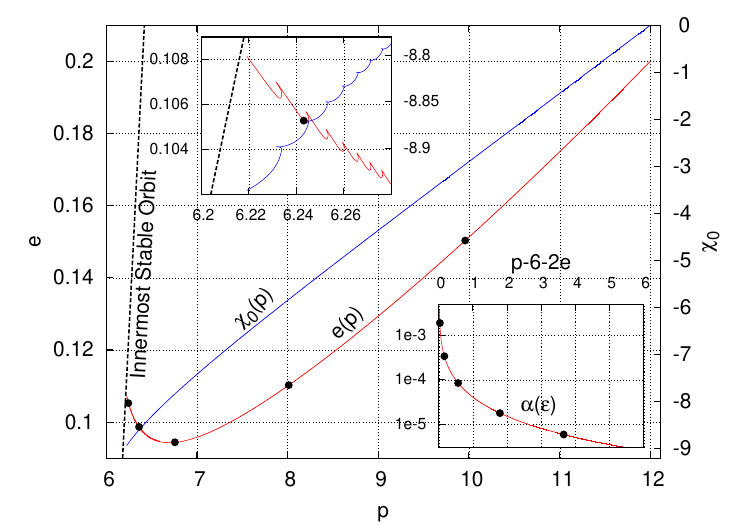}
	\caption{\footnotesize Evolution of the osculating elements in the sample case of Fig.\ \ref{fig:1}. We show the eccentricity $e$ [lighter (red) line, left axis] and periastron phase $\chi_0$ [darker (blue) line, right axis] as functions of semi-latus rectum $p$, as the binary inspirals from $p_0=12$ down to the ISO (dashed curve). Marks along the curves count down (from right to left) 500 days, 100 days, 10 days, 1 day and 1 hour to the onset of plunge. Note the orbit initially circularizes, but upon approaching the ISO the eccentricity begins to increase. Note also the phase $\chi_0$ decreases monotonically, implying that the conservative GSF acts to reduce the rate of relativistic precession. The upper inset is an enlargement of the near-ISO region; the manifest oscillatory behavior is due to the variation of the GSF with the radial phase. The lower inset shows the magnitude of the adiabaticity parameter $\alpha= \langle|\dot{p}/p|T\rangle$ vs.\ the ISO distance $\epsilon=p-6-2e$, confirming that the evolution is strongly adiabatic until very near the ISO. }
	\label{fig:2}
\end{figure}

{\it Sample results.}
Figures \ref{fig:1} and \ref{fig:2} display results from a full-GSF inspiral with $\eta=10^{-5}$, starting at $(p_0,e_0)=(12,0.2)$ (and taking $M = 10^{6}M_{\odot}$ for concreteness). The orbit decays adiabatically (see the lower inset in Fig.\ \ref{fig:2}) and circularizes gradually, until very close to the ISO where the eccentricity begins to increase---a phenomenon already described in Ref.\ \cite{Apostolatos:1993}. The entire inspiral, from $p_0=12$ to the onset of plunge, lasts $\sim 1443\times M/(10^{6}M_{\odot})$ days, during which the orbit completes 75,550 periastron passages. Note the periastron phase $\chi_0$ shifts secularly in a retrograde sense (in our example, by $\sim 9$ radians over the entire inspiral). This represents a GSF-induced decrease in the rate of periastron advance (cf.\ \cite{Barack:2011ed}).

{\it Radiative approximation.}
To explore the long-term effect of the GSF's conservative piece, let us construct an RA model by setting $F_{\rm cons}^{\alpha}=0$ in the evolution equations (\ref{pdot})--(\ref{chidot}), and additionally replacing the expressions on the right-hand side with their corresponding $t$-averages over an entire radial period of the instantaneous osculating geodesic. 
We ask how well this RA model can capture the full-GSF dynamics. 

As a reference for comparison let us consider the accumulated azimuthal phase $\varphi(t)$.
We denote by $\varphi_{\rm RA/full}$ the values corresponding to the RA/full GSF models, and aim to inspect how the phase difference $\Delta\varphi_{\rm RA}\equiv \varphi_{\rm RA}-\varphi_{\rm full}$ builds up over time. To define $\Delta\varphi_{\rm RA}$ unambiguously we must map cautiously between the initial parameters of the RA and full-GSF models, noting the $O(\eta)$ gauge ambiguity in the values of the parameters $p,e$. A mapping based on ``same $p_0,e_0$ values'' would result in the RA and full orbits possessing different initial frequencies, because the conservative piece of the GSF,  which is accounted for in the full model but not in RA, shifts the frequencies by an amount of $O(\eta)$. This, in turn, would result in a rapid linear-in-time growth of $\Delta\varphi_{\rm RA}$. 

To eliminate this spurious effect we instead match the {\it frequencies} of the initial osculating geodesics, using knowledge of $F^{\alpha}_{\rm cons}$. To achieve this in practice we apply the following procedure. (1) Choose $p_0,e_0$ for the full-GSF orbit (taking $\chi_0=\varphi=0$ at $t=0$). (2) Compute the azimuthal and radial frequencies of the orbit at $t=0$ through $O(\eta)$, including $F^{\alpha}_{\rm cons}$-induced corrections. (3) Find the $p,e$ values of a {\em geodesic} whose frequencies are those found in step 2. (4) Use these $p,e$ as initial values for the RA evolution (starting again with $\chi_0=\varphi=0$ at $t=0$). We explain this procedure in more detail in a follow-up work \cite{inprep}, which further explores the performance of the RA model. Our procedure matches the initial frequencies of the full and RA orbits. This is physically motivated because the frequencies (unlike $p,e$) are invariant characteristics of the orbit. 

\begin{figure}[htb]
	\includegraphics[scale=0.95]{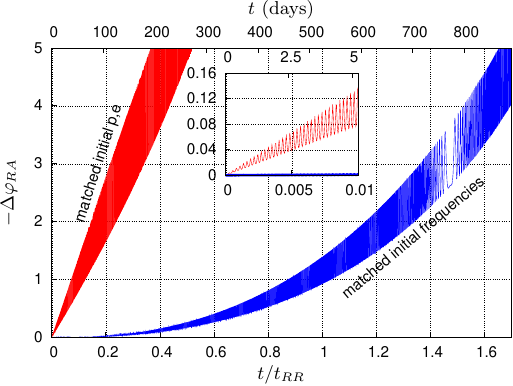}
	\caption{\footnotesize Effect of conservative GSF corrections on the long-term phase evolution. Plotted is the accumulated phase difference $\Delta\varphi_{\rm RA}= \varphi_{\rm RA}-\varphi_{\rm full}$ (RA: Radiative Approximation) for the sample orbit shown in Fig.\ \ref{fig:1}. In the lower (blue) curve we have matched the initial frequencies of the RA and full orbits [correcting for the initial $O(\eta)$ frequency shift due to the conservative piece of the GSF]. The upper (red) curve shows, for reference, the phase difference $\Delta\varphi_{\rm RA}$ when no such adjustment is made. 
}
	\label{fig:3}
\end{figure}

Figure \ref{fig:3} shows $\Delta\varphi_{\rm RA}(t)$ for our sample orbit with $\eta=10^{-5}$ and $(p_0,e_0)=(12,0.2)$. On the lower horizontal axis we express $t$ in units of the radiation-reaction timescale $t_{\rm RR}\equiv T_c/\eta$, where as a characteristic orbital period we take the $\varphi$-period of the innermost stable circular orbit, $T_c=2\pi 6^{3/2}M$.  
As expected, $\Delta\varphi_{\rm RA}$ grows secularly in proportion to $(t/t_{\rm RR})^2$ (with oscillations reflecting the mismatch in radial phase between the RA and full-GSF orbits). This secular growth is attributed to conservative corrections to the rate-of-change of the azimuthal frequency, which are $O(\eta^2)$. The phase difference $\Delta\varphi_{\rm RA}$ remains small for quite long, becoming significant only on a timescale $t\sim t_{\rm RR}$. For reference, we also show in Fig.\ \ref{fig:3} the phase difference {\it without} adjusting the initial frequencies, i.e., using the same $p_0,e_0$ values for both models. In this case $\Delta\varphi_{\rm RA}\propto t$, and $\varphi_{\rm RA}$ quickly drifts away with respect to $\varphi_{\rm full}$. 

The RA model appears to capture the full-GSF phase evolution rather well over an extended portion of the inspiral (allowing for a suitable adjustment of the initial parameters). Our results confirm the expectation that RA-based waveform templates could be implemented usefully in semi-coherent matched filtering searches for gravitational waves from I/EMRIs \cite{Gair:2004iv,Hughes:2005qb,Huerta:2008gb}. To obtain a fully phase-coherent theoretical model of the evolution beyond the radiation-reaction timescale requires the conservative GSF, but it also requires the (as yet unknown) second-order piece of the dissipative GSF. Secular geodesic effects associated with the spin of the small object may also be important at this order. 

{\it Concluding remarks.}
We reported here the development of a computational framework for calculating fully self-forced inspiral orbits for I/EMRI applications. This framework is under continuing development, and several important improvements must be made before we can compute fully coherent waveforms for astrophysical I/EMRIs. First, more accurate GSF data must be obtained and implemented to inform a more accurate interpolation formula; a fractional accuracy of $\lesssim O(\eta)$ in the GSF is desired to ensure that the GSF model error has a negligible long-term effect. Such an accuracy standard is achievable in principle using current codes \cite{Akcay:2010dx,WAB}, but advanced computational techniques now being developed will allow crucial saving in computational cost. Second, our GSF model must include 2nd-order dissipative corrections. Work to formulate and compute such corrections is under way. Third, the model must be extended to Kerr geometry. Techniques to compute the GSF in Kerr are under active development \cite{Warburton:2011hp,Dolan:2011dx}. Once Kerr GSF data are at hand, an interpolation model akin to (\ref{fit}) will need to be devised and implemented. Finally, it remains to quantify the error coming from calculating the GSF along fixed geodesics (and not the true evolving orbit). This must await the completion of a fully self-consistent evolution code, also under development. 

It would be instructive to compare our inspiral orbits with results from fully nonlinear numerical simulations once these become available for small $\eta$. The state of the art is a short simulation for $\eta=1{:}100$ \cite{Lousto:2010ut}, covering the last few orbits of inspiral. A much longer simulation would be required to allow a meaningful comparison with our GSF results.

{\it Acknowledgements.} NW's work is supported by STFC through a studentship grant. SA and LB acknowledge additional support from STFC through grant number PP/E001025/1. JG's work is supported by the Royal Society.   NS acknowledges support by the Grant-in-Aid for Scientific Research (No. 21244033).


\begin{thebibliography}{99}
\bibitem{Mino:1996nk}
  Y.~Mino, M.~Sasaki and T.~Tanaka,
  Phys.\ Rev.\  D {\bf 55}, 3457 (1997).
\bibitem{Poisson:2003nc}
  E.~Poisson,
  Living Rev.\ Rel.\  {\bf 7}, 6 (2004).
\bibitem{Gralla:2008fg}
  S.~E.~Gralla and R.~M.~Wald,
  Class.\ Quant.\ Grav.\  {\bf 25}, 205009 (2008).
\bibitem{Pound:2009sm}
  A.~Pound,
  Phys.\ Rev.\ D {\bf 81}, 024023 (2010).
\bibitem{Brown:2006pj}
  D.~A.~Brown, J. ~Brink, H.~Fang, J.~R.~Gair, C.~Li, G.~Lovelace, I.~Mandel and K.~S.~Thorne,
  Phys.\ Rev.\ Lett.\  {\bf 99}, 201102 (2007).
\bibitem{Gair:2008bx}
  J.~R.~Gair,
  Class.\ Quant.\ Grav.\  {\bf 26}, 094034 (2009).
\bibitem{Barack:2003fp}
  L.~Barack and C.~Cutler,
  Phys.\ Rev.\  D {\bf 69}, 082005 (2004).
\bibitem{AmaroSeoane:2007aw}
  P.~Amaro-Seoane, J.~R.~Gair, M.~Freitag, M.~Coleman Miller, I.~Mandel, C.~J.~Cutler and S.~Babak,
  Class.\ Quant.\ Grav.\  {\bf 24}, R113 (2007).
\bibitem{Barack:2009ux}
  L.~Barack,
  Class.\ Quant.\ Grav.\  {\bf 26}, 213001 (2009).
\bibitem{Barack:2010tm}
  L.~Barack and N.~Sago,
  Phys.\ Rev.\  D {\bf 81}, 084021 (2010).
\bibitem{Barack:2011ed}
  L.~Barack and N.~Sago,
  Phys.\ Rev.\  D {\bf 83}, 084023 (2011).
\bibitem{Favata:2010yd}
  M.~Favata,
  Phys.\ Rev.\  D {\bf 83}, 024027 (2011).
\bibitem{Tiec:2011bk}
  A.~Le Tiec, A.~H.~Mroue, L.~Barack, A.~Buonanno, H.~P.~Pfeiffer, N.~Sago and A.~Taracchini,
  Phys.\ Rev.\ Lett.\  {\bf 107},  141101 (2011).
\bibitem{Tiec:2011dp}
  A.~Le Tiec, E.~Barausse and A.~Buonanno,
	Phys.~ Rev.~Lett.~{\bf 131103}, 108 (2012).
\bibitem{Damour:2009sm}
  T.~Damour,
  Phys.\ Rev.\  D {\bf 81}, 024017 (2010).
\bibitem{Barack:2010ny}
  L.~Barack, T.~Damour and N.~Sago,
  Phys.\ Rev.\  D {\bf 82}, 084036 (2010).
\bibitem{Barausse:2011dq}
  E.~Barausse, A.~Buonanno and A.~Le Tiec,
	Phys.~Rev.~D {\bf 85}, 064010 (2012).
\bibitem{Mino:2003yg}
  Y.~Mino,
  Phys.\ Rev.\  D {\bf 67}, 084027 (2003).
\bibitem{Pound:2007th}
  A.~Pound and E.~Poisson,
  Phys.\ Rev.\  D {\bf 77}, 044013 (2008).
\bibitem{Gair:2010iv}
  J.~R.~Gair, E.~E.~Flanagan, S.~Drasco, T.~Hinderer and S.~Babak,
  Phys.\ Rev.\  D {\bf 83}, 044037 (2011).
\bibitem{Cutler:1994pb}
  C.~Cutler, D.~Kennefick and E.~Poisson,
  Phys.\ Rev.\  D {\bf 50}, 3816 (1994).
\bibitem{Akcay:2010dx}
  S.~Akcay,
  Phys.\ Rev.\  D {\bf 83}, 124026 (2011).
\bibitem{WAB}
 S.~Akcay,  N. Warburton and L. Barack,
	Phys.~Rev.~D {\bf 88}, 104009 (2013).
\bibitem{webpage}
``Gravitational self force calculator'' webpage: \\
\href{http://www.nielswarburton.net/lib_Sch_GSF/}{http://www.nielswarburton.net/lib\_Sch\_GSF/}
\bibitem{Apostolatos:1993}
T. A. Apostolatos, D. Kennefick, A. Ori and E. Poisson, Phys. Rev. D, 47, 5376 (1993).
\bibitem{inprep} 
S. Akcay, L. Barack, R. H. Cole, J. R. Gair, N. Sago and N. Warburton, in preparation. 
\bibitem{Gair:2004iv}
  J.~R.~Gair, L.~Barack, T.~Creighton, C.~Cutler, S.~L.~Larson, E.~S.~Phinney and M.~Vallisneri,
  Class.\ Quant.\ Grav.\  {\bf 21}, S1595 (2004).
\bibitem{Hughes:2005qb}
  S.~A.~Hughes, S.~Drasco, E.~E.~Flanagan and J.~Franklin,
  Phys.\ Rev.\ Lett.\  {\bf 94}, 221101 (2005).
\bibitem{Huerta:2008gb}
  E.~A.~Huerta and J.~R.~Gair,
  Phys.\ Rev.\  D {\bf 79}, 084021 (2009)
  [Erratum-ibid.\  D {\bf 84}, 049903 (2011)].
\bibitem{Warburton:2011hp}
  N.~Warburton and L.~Barack,
  Phys.\ Rev.\  D {\bf 83}, 124038 (2011).
\bibitem{Dolan:2011dx}
  S.~R.~Dolan, B.~Wardell and L.~Barack,
  Phys.\ Rev.\  D {\bf 84}, 084001 (2011).
\bibitem{Lousto:2010ut}
  C.~O.~Lousto and Y.~Zlochower,
  Phys.\ Rev.\ Lett.\  {\bf 106}, 041101 (2011).
\end{thebibliography}
\end{document}